\documentclass[12pt]{article}

\usepackage{amsmath,amssymb,amsbsy,amsfonts,latexsym,graphicx}
\usepackage{color,array}
\usepackage{cite}

\numberwithin{equation}{section}

\allowdisplaybreaks

\evensidemargin \oddsidemargin

\begin{document}



\renewcommand{\thefootnote}{\fnsymbol{footnote}}


\begin{flushright}
KIAS-P14023\\
KUNS-2480
\end{flushright}

\vspace{5mm}
\begin{center}
{\Large\bf No pair production of open strings \\ 
in a plane-wave background
}
\end{center}

\vspace{7mm}
\begin{center}
{\large 
Makoto Sakaguchi$^a$\footnote{\tt msakaguc@mx.ibaraki.ac.jp},
Hyeonjoon Shin$^b$\footnote{\tt hyeonjoon@kias.re.kr} and
Kentaroh Yoshida$^c$\footnote{\tt kyoshida@gauge.scphys.kyoto-u.ac.jp}}
\\[10mm]
 {\sl $^a$Department of Physics, Ibaraki University, 
Mito 310-8512, Japan}
\\[3mm]
{\sl $^b$School of Physics, Korea Institute for Advanced Study\\
  Seoul 130-722, South Korea }
\\[3mm]
{\sl $^c$Department of Physics, Kyoto University\\
Kyoto 606-8502, Japan}

\end{center}

\thispagestyle{empty}

\vspace*{0.1cm}
\begin{center}
{\bf Abstract}
\end{center}
\noindent 
We consider whether an external electric field may cause the pair production of open strings 
in a type IIA plane-wave background. The boundary states of D-branes with condensates 
are constructed in the Green-Schwarz formulation of superstring theory
with the light-cone gauge. 
The cylinder diagrams are computed with massive theta functions. 
Although the value of the electric field is bounded by the upper value,   
there is no pole in the amplitudes and it indicates that 
no pair production occurs in the plane-wave background. 
This result would be universal for a class of plane-wave backgrounds. 

\vfill

\noindent 
Keywords : string theory, plane-wave background, boundary states, pair creation

\vspace{5mm}

\renewcommand{\thefootnote}{\arabic{footnote}}
\setcounter{footnote}{0}

\section{Introduction}

The AdS/CFT correspondence \cite{M,GKP,W} gives a nice laboratory to argue 
new aspects of string theory and gauge theories. The classical dynamics of strings on the AdS space 
is related to nonperturbative phenomena in strongly coupled field theories. 
An example of the fascinating subjects is the pair production of pairs of particle and antiparticle 
in the presence of an external electric field. It is originally studied in quantum electrodynamics (QED) \cite{HE} 
and often called the Schwinger effect \cite{Schwinger} (for further developments, see \cite{AAM,AM}).  

\medskip 

Recently, a holographic scenario to study the Schwinger pair production 
was proposed by Semenoff and Zarembo \cite{SZ}, where the production 
rate of ``W bosons'' in the Coulomb phase 
(for various generalizations, see \cite{AmMa,BKR,SY}). 
The potential analysis is done in the holographic way \cite{SY2}.    
The holographic analysis is applicable to confining gauge theories \cite{SY3,SY4,KSY,Hashi}. 

\medskip 

In relation to this progress, an interesting issue is to consider 
the pair production of strings on the AdS space 
and argue its holographic dual. However, it is technically difficult to analyze 
the dynamics of strings on the AdS space; hence, it would be reasonable to consider a plane-wave background 
as an approximation of the target space geometry. It is known that 
the pair production of strings occurs in flat space \cite{max1,max2} (for the analysis based on 
the Green-Schwarz formulation, see \cite{Tseytlin}), 
but it is not obvious whether it can occur on curved backgrounds. 

\medskip

Here we will focus on a type IIA pp-wave background as a concrete example. 
We consider if an external electric field can induce the pair production of open strings. 
The boundary states of D-branes with condensates are considered in the Green-Schwarz 
formulation of superstring theory with the light-cone gauge. 
The cylinder diagrams are computed with massive theta functions. 
The value of the electric field is found to be bounded by the 
upper value.
On the other hand, there is no pole in the amplitudes and 
it indicates that no pair production occurs in the plane-wave 
background.\footnote{In fact, it has been reported that there is no
vacuum polarization in plane waves in the context of quantum field theory  
\cite{Deser:1875zz}.  Our result may be considered as its stringy 
version.}
Although just one example of plane-wave backgrounds is studied here, 
it is expected that this result would be universal for general plane-wave backgrounds.  

\medskip 

This paper is organized as follows. Section 2 gives a brief review of the Green-Schwarz formulation 
of type IIA superstring theory on a plane-wave background preserving 24 supersymmetries. 
In section 3, boundary states with condensates are constructed. 
In section 4, we compute cylinder amplitudes between parallel D-branes with condensates.  
Section 5 argues the pole structure of the amplitudes after moving to the open string picture. 
It is shown that there is no pole in the amplitudes and therefore 
the pair production does not occur. 
This result indicates that there is no pair production on general plane-wave backgrounds. 
Section 6 is devoted to the conclusion and discussion. 
Appendix A explains in detail that there is no pole in the D2-brane amplitude. 

\section{Setup}

This section gives a brief review of the construction of a light-cone
Hamiltonian for a closed superstring in a type IIA plane-wave background, 
the details of which can be found in \cite{Hyun:2002wp}.

\subsubsection*{A type IIA plane-wave background}

The type IIA plane-wave background is given by \cite{Bena:2002kq,Sugiyama:2002tf,Hyun:2002wu}
\begin{eqnarray}
& & ds^2 = - 2 dx^+ dx^-
    - A(x^I) (dx^+)^2  + \sum^8_{I=1} (dx^I)^2~,
                                      \nonumber \\
& & F_{+123} = \mu~,\ \ \  F_{+4} = -\frac{\mu}{3}~,
\label{pp-wave}
\end{eqnarray}
where $x^I =(x^i,x^{i'})$ and the scalar function $A(x^I)$ is defined as 
\begin{equation}
A(x^I) \equiv \sum^4_{i=1} \frac{\mu^2}{9} (x^i)^2
            +\sum^8_{i'=5} \frac{\mu^2}{36} (x^{i'})^2 \qquad (\mu:~\mbox{const.})\,.
\end{equation}
The relative coefficients are fixed so that the background preserves 24 supersymmetries. 

\subsubsection*{The Green-Schwarz action of type IIA plane-wave string}

The Green-Schwarz action of a type IIA superstring on this background is simplified 
by fixing the kappa symmetry with the light-cone gauge condition, 
\begin{equation}
\Gamma^+ \theta = 0\,, \quad 
X^+ = p^+  \tau\,.
\label{kfix}
\end{equation}
Here $p^+$ is the total momentum conjugate to $X^-$ and $\tau$ is 
time on the world sheet. Then the light-cone gauge fixed
action is  given by\footnote{We set $2 \pi \alpha' = 1$ for our 
convention. Here $\eta^{mn}$ is the flat world-sheet metric
with the world sheet coordinates $\sigma^m=(\tau,\sigma)$\,.}
\begin{eqnarray}
&&S_{LC}
 = - \frac{1}{2} \int  d^2 \sigma
 \Bigg[ \eta^{mn} \partial_m X^I \partial_n X^I
      + \frac{m^2}{9} (X^i)^2
      + \frac{m^2}{36} (X^{i'})^2
                       \nonumber \\
 & & \hspace*{3.2cm} + \bar{\theta} \Gamma^- \partial_\tau \theta
     + \bar{\theta} \Gamma^{-9} \partial_\sigma \theta
     - \frac{m}{4} \bar{\theta} \Gamma^-
        \left( \Gamma^{123} + \frac{1}{3} \Gamma^{49} \right)
        \theta
  \Bigg]\,,
\label{lcaction0}
\end{eqnarray}
where a new parameter $m$ is defined as 
\begin{equation}
m \equiv \mu p^+\,. 
\end{equation}
This is the mass parameter characterizing the masses of the fields on the world sheet. 
Then the Majorana fermion $\theta$ is the combination of
Majorana-Weyl fermions $\theta^1$ and $\theta^2$ with opposite 
ten-dimensional $SO(1,9)$ chiralities; that is, $\theta = \theta^1 +
\theta^2$\,, where $\theta^1~(\theta^2)$ has positive (negative) chirality. 
The Dirac conjugate of $\theta$ is defined as 
$\bar \theta\equiv i \theta^T \Gamma^0$. 

\medskip 

The fermionic part in the action (\ref{lcaction0}) is written in 
the 32-component notation. It is now convenient to
rewrite the action in the 16-component spinor notation.  Let us first
introduce a representation of $SO(1,9)$ gamma matrices as follows: 
\[
 \Gamma^0= -i\sigma^2 \otimes {\bf 1}_{16}\,, \qquad 
 \Gamma^{11}= \sigma^1 \otimes {\bf 1}_{16}\,, \qquad 
 \Gamma^I= \sigma^3 \otimes \gamma^I\,, 
\]
\begin{equation}
 \Gamma^9= -\sigma^3 \otimes \gamma^9\,, \qquad 
 \Gamma^{\pm} = \frac{1}{\sqrt{2}}(\Gamma^0 \pm \Gamma^{11})\,. 
\label{gamma}
\end{equation}
Here $\sigma$'s are Pauli matrices and ${\bf 1}_{16}$ the $16 \times
16$ unit matrix. Then $\gamma^I$ are the $16 \times 16$ symmetric real
gamma matrices satisfying the spin$(8)$ Clifford algebra $\{ \gamma^I,
\gamma^J \} = 2 \delta^{IJ}$, which are reducible to the ${\bf
  8_s}+{\bf 8_c}$ representation of spin$(8)$\,. Note that, in our
representation, $\Gamma^9$ is taken to be the $SO(1,9)$ chirality 
operator and $\gamma^9$ becomes the $SO(8)$ chirality operator,
\begin{equation}
\gamma^9 = \gamma^1 \gamma^2 \cdots \gamma^8\,.
\end{equation}

\medskip 

It is convenient to introduce the spinor notation 
\[
\theta^A = \frac{1}{2^{1/4}} \left(
\begin{array}{c} 0 \\ \psi^A
\end{array} \right)\,, 
\] 
so as to satisfy the kappa symmetry fixing condition of (\ref{kfix})\,, 
where the superscript $A$ denotes the $SO(1,9)$ chirality.  
Then the action $S_{\rm LC}$ is rewritten as 
\begin{eqnarray}
&& S_{\rm LC}
 =  - \frac{1}{2} \int  d^2 \sigma
 \Bigg[\, \eta^{mn} \partial_m X^I \partial_n X^I
      + \frac{m^2}{9} (X^i)^2
      + \frac{m^2}{36} (X^{i'})^2
     - i \psi_+^1 \partial_+  \psi^1_+
     - i \psi_-^1 \partial_+  \psi^1_-  \nonumber \\ 
     && \hspace*{3.5cm}
     - i \psi^2_+ \partial_- \psi^2_+
     - i \psi^2_- \partial_- \psi^2_-
     +2i \frac{m}{3} \psi^2_+ \gamma^4 \psi^1_-
     - 2i \frac{m}{6} \psi^2_- \gamma^4 \psi^1_+
 \Bigg]\,, 
\label{lc-action}
\end{eqnarray}
where we have introduced the light-cone coordinates on the world sheet and the associated derivatives are defined as 
$\partial_\pm \equiv \partial_\tau\pm \partial_\sigma$\,.  
Here the subscripts $\pm$ in $\psi^A_\pm$ represent the eigenvalues $\pm 1$ of
$\gamma^{1234}$\,.  In our convention, the fermion has the same $SO(1,9)$
and $SO(8)$ chiralities measured by $\Gamma^9$ and $\gamma^9$,
respectively.

\subsubsection*{Mode expansions~~(bosons)}

The light-cone action (\ref{lc-action}) is quadratic in fields; hence,
the quantization of closed string is carried out exactly. 

\medskip 

Let us first consider the bosonic sector of the theory.  The equations of
motion for the bosonic coordinates $X^I$ are obtained from the action
(\ref{lc-action}) like 
\begin{equation}
\eta^{mn} \partial_m \partial_n X^i
  - \left( \frac{m}{3} \right)^2 X^i = 0\,, \qquad 
\eta^{mn} \partial_m \partial_n X^{i'}
  - \left( \frac{m}{6} \right)^2 X^{i'} = 0\,.
\label{beom}
\end{equation}
Here the fields are subject to the periodic boundary condition
\[
X^I (\tau, \sigma+2 \pi) = X^I (\tau, \sigma)\,.
\]  
The solutions are given in the form of mode expansion,
\begin{eqnarray}
X^i (\tau,\sigma)
  &=& x^i \cos \left( \frac{m}{3} \tau \right)
     + \frac{1}{2\pi} p^i \frac{3}{m}
             \sin \left( \frac{m}{3} \tau \right) \nonumber \\ 
&& \qquad \qquad   + i \sqrt{ \frac{1}{4\pi} } \sum_{n \neq 0}
         \frac{1}{\omega_n}
      ( \alpha^i_n \phi_n (\tau, \sigma )
    + \tilde{\alpha}^i_n \tilde{\phi}_n (\tau, \sigma) ) \,,
                                   \nonumber \\
X^{i'} (\tau,\sigma)
  &=& x^{i'} \cos \left( \frac{m}{6} \tau \right)
     + \frac{1}{2\pi} p^{i'} \frac{6}{m}
              \sin \left( \frac{m}{6} \tau \right) \nonumber \\ 
&& \qquad \qquad   + i \sqrt{ \frac{1}{4\pi} } \sum_{n \neq 0}
         \frac{1}{\omega'_n}
      ( \alpha^{i'}_n \phi'_n (\tau, \sigma)
     + \tilde{\alpha}^{i'}_n \tilde{\phi}'_n (\tau, \sigma) )\,,
\label{bmode}
\end{eqnarray}
where $x^I$ and $p^I$ are center-of-mass variables, 
coefficients for zero modes, and $\alpha^I_n$ and
$\tilde{\alpha}^I_n$ are the expansion coefficients for the nonzero
modes. The basis functions for nonzero modes are given by
\begin{eqnarray}
& & \phi_n (\tau, \sigma) = {\rm e}^{-i \omega_n \tau - i n \sigma}\,, \qquad 
\tilde{\phi}_n (\tau, \sigma) = {\rm e}^{-i \omega_n \tau + i n \sigma}\,,
\label{mode}  \\
& & \phi'_n (\tau, \sigma) = {\rm e}^{-i \omega'_n \tau - i n \sigma}\,, \qquad 
\tilde{\phi}'_n (\tau, \sigma)
  = {\rm e}^{-i \omega'_n \tau + i n \sigma}\,,
\label{pmode}
\end{eqnarray}
with the wave frequencies
\begin{equation}
\omega_n = {\rm sign}(n)
      \sqrt{ \left( \frac{m}{3} \right)^2 + n^2 }\,, \qquad 
\omega'_n = {\rm sign}(n)
      \sqrt{ \left( \frac{m}{6} \right)^2 + n^2 }\,.
\end{equation}
Note that the reality of $X^I$ requires that $\alpha^{I \dagger}_n
= \alpha^I_{-n}$ and $\tilde{\alpha}^{I \dagger}_n =
\tilde{\alpha}^I_{-n}$\,.

\medskip 

The next is to promote the expansion coefficients in (\ref{bmode}) to operators with the canonical quantization.  
The canonical commutation relations (at equal time) for the bosonic fields are given by 
\begin{equation}
[ X^I (\tau, \sigma), {\cal P}^J (\tau, \sigma') ]
= i \delta^{IJ} \delta (\sigma - \sigma')\,,
\label{bcom}
\end{equation}
where ${\cal P}^J = \partial_\tau X^J$ is the
canonical conjugate momentum of $X^J$\,. Then one can read off the following
commutation relations between the mode operators, 
\begin{eqnarray}
&&[ x^I, p^J ] = i \delta^{IJ}\,, \qquad 
[ \alpha^i_n, \alpha^j_m ] = \omega_n \delta^{ij} \delta_{n+m,0}\,, \qquad 
[ \alpha^{i'}_n, \alpha^{j'}_m ] =
   \omega'_n \delta^{i' j'} \delta_{n+m,0}\, ,
\nonumber \\
&& [ \tilde\alpha^i_n, \tilde\alpha^j_m ] = 
\omega_n \delta^{ij} \delta_{n+m,0}\,, \qquad 
[ \tilde\alpha^{i'}_n, \tilde\alpha^{j'}_m ] =
   \omega'_n \delta^{i' j'} \delta_{n+m,0}\, .
\label{bmcom}
\end{eqnarray}
These relations will be used in considering boundary states in the next section. 

\subsubsection*{Mode expansions~~(fermions)}

For the fermionic sector of the theory, the fermionic fields are split into the two parts like 
$(\psi^1_-, \psi^2_+)$ and $(\psi^1_+, \psi^2_-)$. 
The equations of motion for the former part are obtained as
\begin{equation}
\partial_+ \psi^1_- + \frac{m}{3} \gamma^4 \psi^2_+ = 0\,, \qquad 
\partial_- \psi^2_+ - \frac{m}{3} \gamma^4 \psi^1_- = 0\,.
\label{feom}
\end{equation}
The nonzero mode solutions of these equations are given by using the
modes, (\ref{mode}). For the zero-mode part of the solution, we impose
a condition that, at $\tau=0$, the solution behaves just as that of the
massless case.  The mode expansions for the fermionic coordinates are
then
\begin{eqnarray}
\psi^1_- (\tau, \sigma)
 &=& c_0 \tilde{\psi}_0 \cos \left( \frac{m}{3} \tau \right)
     - c_0 \gamma^4 \psi_0 \sin \left( \frac{m}{3} \tau \right)
                              \nonumber \\
 & & \qquad  + \sum_{n \neq 0} c_n
      \left(
         \tilde{\psi}_n \tilde{\phi}_n (\tau, \sigma)
         - i \frac{3}{m} (\omega_n - n)
            \gamma^4 \psi_n \phi_n (\tau, \sigma )
      \right)\,,
                               \nonumber \\
\psi^2_+ (\tau, \sigma)
 &=& c_0 \psi_0 \cos \left( \frac{m}{3} \tau \right)
     + c_0 \gamma^4 \tilde{\psi}_0
        \sin \left( \frac{m}{3} \tau \right)
                              \nonumber \\
 & & \qquad + \sum_{n \neq 0} c_n
      \left(
         \psi_n \phi_n (\tau, \sigma)
         + i \frac{3}{m} (\omega_n - n)
            \gamma^4 \tilde{\psi}_n \tilde{\phi}_n (\tau, \sigma )
      \right)\,,
\label{fmode}
\end{eqnarray}
where the chirality condition is that $\gamma^{1234} \psi_n = \psi_n$ and $\gamma^{1234}
\tilde{\psi}_n = -\tilde{\psi}_n$ for all $n$\,. The normalization constants $c_0$ and $c_n$ are given by
\[
c_0 = \frac{1}{\sqrt{2\pi}}\,, \qquad 
c_n = \frac{1}{\sqrt{2\pi}}\frac{1}{\sqrt{1 +
          \left( \frac{3}{m} \right)^2 (\omega_n - n)^2 } }\,.
\]

\medskip 

Then let us consider the other part $(\psi^1_+, \psi^2_-)$\,. 
The equations of motion are given by, respectively, 
\begin{equation}
\partial_+ \psi^1_+ - \frac{m}{6} \gamma^4 \psi^2_- = 0\,, \qquad 
\partial_- \psi^2_- + \frac{m}{6} \gamma^4 \psi^1_+ = 0\,.
\label{fpeom}
\end{equation}
The solutions are found to be
\begin{eqnarray}
\psi^1_+ (\tau, \sigma)
 &=& c'_0 \tilde{\psi}'_0 \cos \left( \frac{m}{6} \tau \right)
     + c'_0 \gamma^4 \psi'_0 \sin \left( \frac{m}{6} \tau \right)
                              \nonumber \\
 & & \qquad + \sum_{n \neq 0} c'_n
      \left(
         \tilde{\psi}'_n \tilde{\phi}'_n (\tau, \sigma)
         + i \frac{6}{m} (\omega'_n - n)
            \gamma^4 \psi'_n \phi'_n (\tau, \sigma )
      \right)\,,
                               \nonumber \\
\psi^2_- (\tau, \sigma)
 &=& c'_0 \psi'_0 \cos \left( \frac{m}{6} \tau \right)
     - c'_0 \gamma^4 \tilde{\psi}'_0
        \sin \left( \frac{m}{6} \tau \right)
                              \nonumber \\
 & & \qquad  + \sum_{n \neq 0} c'_n
      \left(
         \psi'_n \phi'_n (\tau, \sigma)
         - i \frac{6}{m} (\omega'_n - n)
            \gamma^4 \tilde{\psi}'_n \tilde{\phi}'_n (\tau, \sigma )
      \right)\,,
\label{fpmode}
\end{eqnarray}
where the chirality conditions are described by $\gamma^{1234} \psi'_n = -\psi'_n$ and $\gamma^{1234}
\tilde{\psi}'_n = \tilde{\psi}'_n$\,. The normalization constants $c_0$ and $c_n$ are given by 
\[
c'_0 = \frac{1}{\sqrt{2\pi}}\,, \qquad 
c'_n = \frac{1}{\sqrt{2\pi}} \frac{1}{\sqrt{1 +
          \left( \frac{6}{m} \right)^2 (\omega'_n - n)^2 } }\,.
\]

\medskip 

Promoting the expansion coefficients to operators with the canonical quantization again, 
the canonical anticommutation relations (at equal time) are obtained as 
\begin{equation}
\{ \psi^A_\pm (\tau, \sigma), \psi^B_\pm (\tau, \sigma') \}
= \delta^{AB} \delta (\sigma - \sigma')\,. 
\label{fcom}
\end{equation}
Then the anticommutation relations between the mode operators are determined as 
\begin{eqnarray}
&& \{ \psi_n , \psi_m \} = \delta_{n+m,0}\,, \qquad 
\{ \tilde{\psi}_n, \tilde{\psi}_m \} = \delta_{n+m,0}\,,
\label{fmcom} \\ 
&& \{ \psi'_n , \psi'_m \} = \delta_{n+m,0}\,, \qquad 
\{ \tilde{\psi}'_n, \tilde{\psi}'_m \} = \delta_{n+m,0}\,.
\label{fpmcom}
\end{eqnarray}

\subsubsection*{The light-cone Hamiltonian}

The light-cone Hamiltonian of the theory is given by  
\begin{equation}
H = \frac{1}{p^+} \int^{2\pi}_0\! d \sigma\, {\cal H}\,, 
\label{lc-ham}
\end{equation}
with the Hamiltonian density ${\cal H}$ derived from $S_{\rm LC}$ given in (\ref{lc-action}),
\begin{eqnarray}
{\cal H}
 &=&  \frac{1}{2} ({\cal P}^I)^2
    + \frac{1}{2} (\partial_\sigma X^I)^2
    +\frac{1}{2} \left( \frac{m}{3} \right)^2 (X^i)^2
    +\frac{1}{2} \left( \frac{m}{6} \right)^2 (X^{i'})^2
  \nonumber \\
 & & - \frac{i}{2} \psi^1_- \partial_\sigma \psi^1_-
     + \frac{i}{2} \psi^2_+ \partial_\sigma \psi^2_+
     + i \frac{m}{3} \psi^2_+ \gamma^4 \psi^1_-
  \nonumber \\
 & & - \frac{i}{2} \psi^1_+ \partial_\sigma \psi^1_+
     + \frac{i}{2} \psi^2_- \partial_\sigma \psi^2_-
     - i \frac{m}{6} \psi^2_- \gamma^4 \psi^1_+ ~.
\end{eqnarray}
By plugging the mode expansions for the fields, Eqs.~(\ref{bmode}),
(\ref{fmode}), and (\ref{fpmode}), into Eq.~(\ref{lc-ham}), the
light-cone Hamiltonian becomes
\begin{equation}
H = E_0 + E + \tilde{E}\,,
\label{lc-h}
\end{equation}
where $E_0$\,, $E$, and $\tilde{E}$ are given by 
\begin{eqnarray}
E_0 &=& \frac{\pi}{p^+}
  \left[ \left( \frac{p^I}{2\pi} \right)^2
       + \left( \frac{m}{3} \right)^2 (x^i)^2
       + \left( \frac{m}{6} \right)^2 (x^{i'})^2
       - \frac{i}{\pi} \frac{m}{3} \tilde{\psi}_0 \gamma^4 \psi_0
       + \frac{i}{\pi} \frac{m}{6} \tilde{\psi}'_0 \gamma^4 \psi'_0
  \right]\,,
  \nonumber \\
E &=& \frac{1}{p^+} \sum^\infty_{n =1}
  ( \alpha^I_{-n} \alpha^I_n
   + \omega_n \psi_{-n} \psi_n
   + \omega'_n \psi'_{-n} \psi'_n
  )\,,
  \nonumber \\
\tilde{E}
  &=& \frac{1}{p^+} \sum^\infty_{n =1}
  ( \tilde{\alpha}^I_{-n} \tilde{\alpha}^I_n
   + \omega_n \tilde{\psi}_{-n} \tilde{\psi}_n
   + \omega'_n \tilde{\psi}'_{-n} \tilde{\psi}'_n
  )\,.
\label{nzero-h}
\end{eqnarray}
The first part $E_0$ is the zero-mode contribution and the remaining two parts, $E$ and $\tilde{E}$\,, are the
contributions of the nonzero modes. 
Here $E$ and $\tilde{E}$ are the normal-ordered expressions, which
do not have zero-point energy in total because  the zero-point energy coming 
from the bosonic fields is exactly canceled by that of the fermionic ones.

\medskip 

The zero-mode contribution $E_0$ has the form of simple harmonic 
oscillators. Hence it can be conveniently rewritten in terms of 
the creation and annihilation operators. 
For the bosonic part, the creation and annihilation operators are defined as 
\begin{eqnarray}
& & a^{i \dagger} \equiv \sqrt{ \frac{3 \pi}{m} }
      \left( \frac{p^i}{2 \pi} + i \frac{m}{3} x^i \right)\,,
 \qquad a^i \equiv \sqrt{ \frac{3 \pi}{m} }
      \left( \frac{p^i}{2 \pi} - i \frac{m}{3} x^i \right)\,,
  \nonumber \\
& & a^{i' \dagger} \equiv \sqrt{ \frac{6 \pi}{m} }
    \left( \frac{p^{i'}}{2 \pi} + i \frac{m}{6} x^{i'} \right)\,,
 \qquad a^{i'} \equiv \sqrt{ \frac{6 \pi}{m} }
    \left( \frac{p^{i'}}{2 \pi} - i \frac{m}{6} x^{i'} \right)\,,
\label{bz}
\end{eqnarray}
and for the fermionic part, those are given by 
\begin{eqnarray}
& & \chi^\dagger \equiv \frac{1}{\sqrt{2}}
             ( \psi_0 - i \gamma^4 \tilde{\psi}_0 )\,,
 \qquad \chi \equiv \frac{1}{\sqrt{2}}
             ( \psi_0 + i \gamma^4 \tilde{\psi}_0 )\,,
  \nonumber \\
& & \chi'^\dagger \equiv \frac{1}{\sqrt{2}}
             ( \psi'_0 + i \gamma^4 \tilde{\psi}'_0 )\,,
 \qquad \chi' \equiv \frac{1}{\sqrt{2}}
             ( \psi'_0 - i \gamma^4 \tilde{\psi}'_0 )\,.
\label{fz}
\end{eqnarray}
Note that the chirality conditions are now rewritten as $\gamma^{12349} \chi = - \chi$ and $\gamma^{12349} \chi' =
\chi'$.  Equations (\ref{bmcom}), (\ref{fmcom}) and (\ref{fpmcom}) lead to 
the nonvanishing (anti)commutation relations,  
\begin{equation}
[ a^I, a^{J \dagger} ] = \delta^{IJ} \,, \qquad
\{ \chi, \chi^\dagger \} = 1 \,, \qquad
\{ \chi', \chi'^\dagger \} = 1 \,.
\label{bzalg}
\end{equation}
Then, after taking the normal ordering, the zero-mode contribution to $H$ 
is given by
\begin{equation}
E_0 = \frac{m}{6p^+}
     ( 2 a^{i \dagger} a^i + a^{i' \dagger} a^{i'}
       + 2 \chi^\dagger \chi + \chi'^\dagger \chi' )\,.
\label{zero-h}
\end{equation}
Here the zero-point energy vanishes as in the case
of $E$ and $\tilde{E}$\,. 

\medskip 

The light-cone Hamiltonian $H$ of (\ref{lc-h}) with (\ref{nzero-h}) and
(\ref{zero-h}) will be used to describe time evolution of closed string states.

\section{Boundary states with condensates}
\label{brane}

In this section, boundary states for D-branes with condensates are constructed in 
the Green-Schwarz formulation of type IIA superstring theory 
on the plane-wave background. The boundary states constructed here will be utilized to 
compute the cylinder diagrams between parallel D-branes in the closed-string description. 

\subsubsection*{The bosonic part}

It is well known that, in the 
light-cone gauge (\ref{kfix}), the light-cone closed-string coordinates 
$X^\pm$ satisfy the Dirichlet boundary condition on the boundary state
for D-brane basically due to the open/closed string channel duality \cite{Green:1996um,Billo:2002ff} .
That is, letting the boundary state $|B\rangle$, it follows that 
\[
\partial_\sigma X^\pm |B\rangle = 0\,.
\] 
These conditions imply that D-branes are instantonic objects and
restrict the dimensionality of a D$p$-brane to the range $0 \le p+1 \le 8$\,.

\medskip 

For the spatial coordinates, the boundary condition can be taken as 
\begin{align}
(\partial_+ X^I - M_{IJ} \partial_- X^J) |B\rangle = 0 \,,
\label{bbc}
\end{align}
where $\partial_\pm = \partial_\tau \pm \partial_\sigma$ as defined
in (\ref{lc-action}). The matrix $M_{IJ}$ eventually describes a rotation. 
In the absence of boundary condensates, its explicit
form is given by
\begin{align}
M_{IJ} = \pm \delta_{IJ}  \qquad \quad 
\left\{
\begin{array}{cc}
+ & \text{for}~I \in D \\ 
- & \text{for}~I \in N
\end{array}
\right.\,,
\label{fbbc}
\end{align}
where $I \in D \, (N)$ means that $I$ denotes the Dirichlet (Neumann) 
direction. Plugging this matrix into (\ref{bbc}), the Dirichlet or Neumann boundary condition 
is imposed for $X^I$ like 
\[
\partial_\sigma X^I |B\rangle =0\,, \qquad \mbox{or} \qquad \partial_\tau X^I |B\rangle =0\,.
\]
In the presence of boundary condensates, $M_{IJ}$ is no longer the form of
(\ref{fbbc}) as it should be.

\medskip 

Note that, even without knowing the explicit form of 
$M_{IJ}$ in the presence of boundary condensates, 
boundary states can be constructed at least at the formal level.  In terms of 
the bosonic modes in the mode expansion (\ref{bmode}) with (\ref{bz}), 
the boundary condition (\ref{bbc}) is rewritten as
\begin{align}
& (a^I - M_{IJ} a^{J\dagger} ) | B \rangle =0 \,,
\notag \\
& (\alpha_n^I - M_{IJ} \tilde{\alpha}^J_{-n}) | B \rangle =0 
  \quad \quad (n \ge 1) \,.
\end{align}
Then one can solve these conditions with 
the method of constructing the coherent state. 
The resulting state is given by 
\begin{align}
|B\rangle_{\rm B} = 
{\rm e}^{\sum_{n>0} 
    \left(
        \frac{1}{\omega_n} M_{ij} \alpha_{-n}^i \tilde{\alpha}_{-n}^j
      + \frac{1}{\omega'_n} M_{i'j'} \alpha_{-n}^{i'} 
                                    \tilde{\alpha}_{-n}^{j'}
     \right)
   }
          {\rm e}^{\frac{1}{2} M_{ij} a^{i\dagger} a^{j\dagger} 
            +\frac{1}{2} M_{i'j'} a^{i'\dagger} a^{j'\dagger}}
          | 0 \rangle \,,
\label{bbs}
\end{align}
where the subscript ${\rm B}$ in $|B\rangle_{\rm B}$ means the bosonic part of 
the boundary state.  This expression shows that the problem of 
constructing a boundary state reduces to the problem of finding
an appropriate matrix $M_{IJ}$ under a given setup.

\medskip 

In the following, we are concerned with the nonvanishing boundary condensates. 
Here the condensates are supposed to be constant electromagnetic fields 
$\mathcal{F}_{IJ}$ on the D-brane world volume and hence all the 
indices of $\mathcal{F}_{IJ}$ are in Neumann directions.  

\medskip 

In the presence of $\mathcal{F}_{IJ}$, the boundary condition for Neumann
directions is given by 
\begin{align}
(\partial_\tau X^I + \mathcal{F}_{IJ} \partial_\sigma X^J)|B\rangle = 0 \quad \quad (I,J \in N)\,.
\end{align}
By rewriting this condition in terms of $\partial_\pm$\,, the relation between $M_{IJ}$
and $\mathcal{F}_{IJ}$ is obtained as
\begin{align}
M_{IJ} = - \left(
\frac{1-\mathcal{F}}{1+\mathcal{F}}
\right)_{IJ} \,\qquad (I,J \in N)\,,
\label{f2m}
\end{align}
while $M_{IJ}=\delta_{IJ}$ when $I,J\in D$.
It is convenient here to take a frame so that 
$\mathcal{F}_{IJ}$ becomes the block-diagonal form,
\begin{align}
\mathcal{F} = \text{diag}
( \mathcal{F}_{(1)}, \mathcal{F}_{(2)}, \dots,
\mathcal{F}_{(p/2)} ) \,,
\label{diagf}
\end{align}
where each block $\mathcal{F}_{(a)}$ is given by
\begin{align}
\mathcal{F}_{(a)} = 
\left(
\begin{array}{cc}
 0 & f_a \\ -f_a & 0
\end{array}
\right) \,.
\label{blockf}
\end{align}
From (\ref{f2m}), we see that $M_{IJ}$ has also the block-diagonal form.
Recalling that $M_{IJ}$ is a rotation matrix, the 
$2\times 2$ block $M_{(a)}$ of $M$ associated with $\mathcal{F}_{(a)}$
represents a rotation in a certain two-dimensional plane
labeled by $a$:
\begin{align}
M_{(a)} 
&= -\frac{1}{1+f_a^2}
\left( \begin{array}{cc} 
1-f_a^2 & -2 f_a \\ 2 f_a & 1-f_a^2 
\end{array}
\right)
\notag \\
&= e^{\varphi_a T_{(a)}}
= \left( \begin{array}{cc} 
\cos \varphi_a & \sin \varphi_a 
\\ -\sin \varphi_a & \cos \varphi_a
\end{array}
\right) \,. 
\label{bmblock}
\end{align}
Here the first line is derived from (\ref{f2m}) with (\ref{diagf})
and (\ref{blockf}), $T_{(a)}$ is the rotation generator in the 
two-dimensional plane labeled by $a$,
\begin{align}
T_{(a)}=\left( \begin{array}{cc} 0 & 1 \\ -1 & 0 \end{array} \right) \,,
\label{rotgen}
\end{align}
and $\varphi_a$ is the rotation
angle related to the constant background $f_a$ as follows: 
\begin{align}
\cos \varphi_a = - \frac{1-f_a^2}{1+f_a^2}
\,, \qquad
\sin \varphi_a = \frac{2f_a}{1+f_a^2} \,.
\label{2dangle}
\end{align}

\subsubsection*{The fermionic part}

Next we consider the fermionic part.  The fermionic 
boundary state in the absence of boundary condensates has been
constructed in \cite{Shin:2003ae}.  Like the bosonic part, the
basic structure of the fermionic boundary state is not changed even
in the presence of boundary condensates.  Thus we will just quote
some essential results obtained in \cite{Shin:2003ae} without repeating
the detailed analysis and focus upon the fermionic counterpart of the
matrix $M_{IJ}$\,.

\medskip

The fermionic boundary state is constructed by requiring that the
boundary state preserves some amount of supersymmetries possessed by the 
IIA plane-wave background. Here, we demand that half of the 
supersymmetries are preserved (i.e., 1/2-BPS condition), 
and that the fermionic modes satisfy the following Ans\"{a}tze as the 
boundary conditions,
\begin{align}
&
(\chi-\gamma^{123}\hat M\chi^\dagger)|B\rangle =0 \,,
\quad
(\chi'-\gamma^{123}\hat M\chi'^\dagger)|B\rangle =0 \,,
\notag\\
&(\tilde{\psi}_n + i \hat{M} \psi_{-n})
|B \rangle = 0 \,,
\notag \\
&(\tilde{\psi}'_n + i \hat{M} \psi'_{-n})
|B \rangle = 0   \quad \quad (n > 0) \,.
\label{fansatz}
\end{align}
Here $\hat{M}$ is an orthogonal matrix and the fermionic counterpart 
of the matrix $M_{IJ}$.
The consistency of 1/2 supersymmetry preserving condition with these
Ans\"{a}tze leads to the conditions which determine the matrix $\hat{M}$,
\begin{align}
&\gamma^J M_{JI} + \hat{M} \gamma^I \hat{M}^T= 0 \,,
\notag \\
&M_{JI} \gamma^J \gamma^{123} + 
\hat{M} \gamma^I \gamma^{123} \hat{M} =0 \,.
\label{mhateq}
\end{align}

\medskip 

In the absence of boundary condensates, by using  
$M_{IJ}$ of (\ref{fbbc}), the solution that satisfies the conditions (\ref{mhateq}) 
is simply given by the product of gamma
matrices with indices in a particular set of Neumann directions, 
\begin{align}
\hat{M} =\gamma^{I_1 I_2 \cdots I_{p+1}} \qquad (I_m \in N\,,~~  I_1 < I_2 < \dots <I_{p+1}) \,,
\label{fmhat}
\end{align}  
for each 1/2-BPS D$p$-brane boundary state. 
In general, possible configurations of D$p$-branes are restricted on plane-wave backgrounds. 
The possible configurations in the present case \cite{Shin:2003ae} are summarized in Table \ref{list:tab}. 

\begin{table}
\begin{center}
 \begin{tabular}{c|c}
\hline
$\#_N ~(= p+1)$ &  Spanning directions \\
\hline 
\hline 
1  &  $(\hat{i})$ \\ 
3 & $(\hat{i},\hat{j},4)$ \,, \quad $(4, i', j')$ \\
5 &  $(1,2,3,i',j')$\,, \quad $(\hat{i},5,6,7,8)$ \\
7 & $(\hat{i},\hat{j},4,5,6,7,8)$ \\
\hline
 \end{tabular}
\end{center}
\caption{\footnotesize Spanning directions of D$p$-brane instantons in the type IIA 
plane-wave background. The branes are 1/2-BPS when sitting at the origin of the transverse
space. $\#_N$ denotes the number of Neumann directions. The indices are defined as 
$\hat{i}, \hat{j} = 1,2,3$ and $i', j' = 5,6,7,8$\,.} 
\label{list:tab}
\end{table}

\medskip 

Like in the bosonic case, the fermionic boundary state can be constructed 
even without knowing the explicit form of $\hat{M}$ in the presence of 
boundary condensates. Namely, by solving the Ans\"{a}tze (\ref{fansatz}) 
with the method of constructing the coherent state, 
the fermionic boundary state is obtained as
\begin{align}
|B \rangle_{\rm F} = 
{\rm e}^{-i\sum_{n>0} 
          \left(
             \psi_{-n} \hat{M} \tilde{\psi}_{-n}
            + \psi'_{-n} \hat{M} \tilde{\psi}'_{-n}
          \right)}
{\rm e}^{\frac{1}{2} \chi^\dagger \gamma^{123} \hat{M} \chi^\dagger 
+\frac{1}{2} \chi'^\dagger \gamma^{123} \hat{M} \chi'^\dagger} 
|0\rangle \,,
\label{fbs}
\end{align}
where the subscript ${\rm F}$ in $|B\rangle_{\rm F}$ means the fermionic part of 
the boundary state and the redefined fermionic zero modes $\chi$ and $\chi'$ of (\ref{fz}) have been used.

\medskip

Note that $\hat{M}$ is related to $M_{IJ}$ under the conditions given in (\ref{mhateq})\,. Hence,
when boundary condensates in the form of (\ref{diagf}) are turned on, 
$\hat{M}$ is determined from $M_{IJ}$ of (\ref{f2m}) with the $2\times 2$ blocks given in (\ref{bmblock}).
It should be remarked that (\ref{f2m}) is only for the Neumann directions but
$M_{IJ}$ for the Dirichlet directions is still given by (\ref{fbbc}). 
Though one may directly solve (\ref{mhateq}) for $\hat{M}$, it is
easier to determine $\hat{M}$ by taking the group theoretical
viewpoint that $\hat{M}$ and $M$ are the rotation matrices
in the spinor and vector representations respectively.  

\medskip 

This can be illustrated by an explicit example. 
Let us consider a D2-brane spanning along the 1, 2 and 4 directions (see Table \ref{list:tab}) 
and turn on a boundary condensate on the 1-2 plane.
Then, $M_{IJ}$ is given by $({\rm e}^{\varphi T})_{IJ}$ for $I,J=1,2$\,,
and otherwise by (\ref{fbbc}). Here, $T$ is the rotation generator in the 1-2 plane given 
by (\ref{rotgen}), and $\varphi$ is related to the boundary condensate through (\ref{2dangle})\footnote{
The subscript in (\ref{2dangle}), which 
distinguishes two-dimensional planes, is not necessary in the present
case.}.  Now, the spinor representation of the rotation in the 1-2 plane by
an angle $\varphi$ is simply given by ${\rm e}^{\frac{\varphi}{2} \gamma^{12}}$,
which is the part of $\hat{M}$ that realizes the rotation. 
The full form of $\hat{M}$ is given by 
\begin{align}
\hat{M} 
= {\rm e}^{\frac{\varphi}{2} \gamma^{12}} \gamma^4 
= \left( \cos \frac{\varphi}{2} +
         \gamma^{12} \sin \frac{\varphi}{2}
  \right) \gamma^4 \,,
\end{align}
which satisfies (\ref{mhateq}), as it should be. 
Note that the boundary condensate is absent when $\varphi = \pi$\,, 
as can be seen from (\ref{2dangle}). 
Then $\hat{M}$ is reduced to just $\gamma^{124}$, and this is nothing but 
the expression implied by (\ref{fmhat}).

\medskip 

Also for the other branes listed in Table \ref{list:tab}, 
various boundary condensates can be turned on and 
the corresponding $\hat{M}$'s are determined by following the same manner. 
Some examples will be given in the next section.

\subsubsection*{The full boundary state}

In summary, by collecting the bosonic and fermionic parts of the boundary
state, (\ref{bbs}) and (\ref{fbs}), the full boundary state
for a D$p$-brane instanton $| Dp \rangle$ is given by 
\begin{align}
| {\rm D}p \rangle 
 = \mathcal{N}
 {\rm e}^{\sum_{n>0} 
    \left(
        \frac{1}{\omega_n} M_{ij} \alpha_{-n}^i \tilde{\alpha}_{-n}^j
      + \frac{1}{\omega'_n} M_{i'j'} \alpha_{-n}^{i'} 
                                    \tilde{\alpha}_{-n}^{j'}
     -i \psi_{-n} \hat{M} \tilde{\psi}_{-n}
     + i\psi'_{-n} \hat{M} \tilde{\psi}'_{-n}
     \right)
   }
| {\rm D}p \rangle_0 \,,
\end{align}
where $\mathcal{N}$ is the normalization constant and 
$| {\rm D}p \rangle_0$ is the part composed of zero modes,
\begin{align}
| {\rm D}p \rangle_0
= {\rm e}^{\frac{1}{2} M_{ij} a^{i\dagger} a^{j\dagger} 
            +\frac{1}{2} M_{i'j'} a^{i'\dagger} a^{j'\dagger}}
  {\rm e}^{\frac{1}{2} \chi^\dagger \gamma^{123} \hat{M} \chi^\dagger 
     +\frac{1}{2} \chi'^\dagger \gamma^{123} \hat{M} \chi'^\dagger}
          | 0 \rangle \,.
\end{align}
In the next section we will compute cylinder diagrams with the constructed boundary states.

\section{Parallel branes with boundary condensates}

In this section, we compute cylinder diagrams which describe the interaction between two parallel D-branes. 
The typical expression of the amplitude is given by 
\begin{align}
& \mathcal{A}_{{\rm D}p_1;{\rm D}p_2} (X^+,X^-,\mathbf{q}_1,\mathbf{q}_2) 
\notag \\
&= \int \frac{dp^+dp^-}{2\pi i}\, {\rm e}^{-ip^+ X^- - i p^- X^+}
\langle {\rm D}p_1 , -p^+, -p^-, \mathbf{q}_1 | 
  \left( \frac{1}{p^+ (p^- - H)} \right)
| {\rm D}p_2, p^+, p^-, \mathbf{q}_2 \rangle
\notag \\
&= \int^\infty_{-\infty} \frac{dp^+}{p^+}\, {\rm e}^{-ip^+ X^-}\theta(p^+)
\langle {\rm D}p_1 , -p^+, \mathbf{q}_1 | 
 {\rm e}^{-i H X^+}
| {\rm D}p_2, p^+, \mathbf{q}_2 \rangle\,, 
\end{align}
where $H$ is the closed-string light-cone Hamiltonian (\ref{lc-h}) and
 $X^\pm = (x^\pm_2-x^\pm_1)$ are the separation of two
branes in the light-cone directions. Then $\mathbf{q}_1$ and
$\mathbf{q}_2$ describe the transverse positions.  The 
$i\epsilon$ prescription,which induces  
the step function $\theta(p^+)$ \cite{Bergman:2002hv},
 is used in going to the last line.

\medskip 

Let us perform a usual Wick rotation on the world sheet, $t=i\tau/\pi$ 
(with $\pi$ for later convenience), or in terms of the string coordinate, 
\begin{align}
t = i \frac{X^+}{\pi p^+} \,.
\end{align}
Then the resulting amplitude has the form
\begin{align}
\mathcal{A}_{{\rm D}p_1;{\rm D}p_2} (X^+,X^-,\mathbf{q}_1,\mathbf{q}_2)
= \int^\infty_0 \frac{dt}{t}\, {\rm e}^{\frac{X^+X^-}{\pi t}}
\tilde{\mathcal{A}}_{{\rm D}p_1;{\rm D}p_2} (t, \mathbf{q}_1, \mathbf{q}_2) \,,
\end{align}
where we have introduced the following quantity, 
\begin{align}
\tilde{\mathcal{A}}_{{\rm D}p_1;{\rm D}p_2} (t, \mathbf{q}_1,\mathbf{q}_2)
= \langle {\rm D}p_1 , -p^+, \mathbf{q}_1 | {\rm e}^{-2\pi t (Hp^+/2)} |
  {\rm D}p_2, p^+, \mathbf{q}_2 \rangle \,.
\end{align}
In the following, we will consider the amplitude with identical D-branes (i.e., $p_1=p_2$) sitting 
at the origin of the transverse space, that is, 
$\mathbf{q}_1= \mathbf{q}_2 = 0$\,.

\subsection{A general prescription to compute the amplitudes}

The amplitude is evaluated in the standard way, and
all the following building blocks for the amplitude calculation are
obtained by following \cite{Green:1996um,Bergman:2002hv,Gaberdiel:2002hh}.  
Note that the zero-point energy is not taken into account 
because it is canceled out between the bosonic and fermionic contributions 
in the final expression.  

\subsubsection*{The bosonic part (for nonzero modes)}

Let us first see the bosonic oscillator part. 
For each of the Neumann directions without boundary condensate and 
each of the Dirichlet directions, 
when the direction is in the 1234 space (or in the 5678 space), 
the contribution is given by 
\begin{align}
\prod_{n>0}(1-q^{\omega_n})^{-1} \qquad \left(\mbox{or} \quad 
 \prod_{n>0}(1-q^{\omega'_n})^{-1} \right)\,, \qquad q \equiv {\rm e}^{-2\pi t}\,.
\label{builds}
\end{align}

\medskip 

For the presence of boundary condensates in a two-dimensional plane labeled by $a$\,, 
when the two-dimensional plane is in the 1234 space, the contribution is given by 
\begin{align}
\prod_{n>0}(1-q^{\omega_n} {\rm e}^{i \phi_a})^{-1} 
  (1-q^{\omega_n} {\rm e}^{-i \phi_a})^{-1}\,, 
\end{align}
where $\phi_a$ is the difference between two 
boundary condensates on the two parallel branes represented by $\varphi_a^{(1)}$ and $\varphi_a^{(2)}$\,, 
respectively,
\begin{align}
\phi_a = \varphi_a^{(1)} - \varphi_a^{(2)} \,.
\label{phi_a}
\end{align}
From (\ref{2dangle})\,, the angles that describe the boundary condensates are represented by 
\begin{align}
\cos \varphi^{(1)}_a = - \frac{1-f^{(1)2}_a}{1+f^{(1)2}_a}
\,, \qquad
\cos \varphi^{(2)}_a = - \frac{1-f^{(2)2}_a}{1+f^{(2)2}_a} \,.
\end{align}
On the other hand,  when the two-dimensional plane is in the 5678 space, 
the contribution is given by 
\begin{align}
\prod_{n>0}(1-q^{\omega'_n} {\rm e}^{i \phi_a})^{-1}
  (1-q^{\omega'_n} {\rm e}^{-i \phi_a})^{-1}\,. 
\end{align}

\subsubsection*{The fermionic part (for nonzero modes)}

For the fermionic oscillator part, the contribution 
from the modes $\psi_n$ and $\tilde{\psi}_n$ is given by 
\begin{align}
\prod_{n>1} \prod_{s_1,s_2,\dots = \pm 1}
\left( 
    1- q^{\omega_n} {\rm e}^{\frac{i}{2}(s_1 \phi_1+s_2 \phi_2 + \dots)}
\right)^{d(s_1,s_2,\dots)}\,, 
\label{fosc}
\end{align}
where $s_a$'s are the eigenvalues of the spinors for the 
rotation generator given by the product of 
two $\gamma$ matrices in the corresponding two-dimensional plane labeled 
by $a$\,.\footnote{The eigenvalues of the antisymmetric product of two 
$\gamma$ matrices are $\pm i$. The symbol $s_a$ for these
eigenvalues represents only the sign.} 
The difference $\phi_a$ is given in (\ref{phi_a}).
Then $d(s_1,s_2,\dots)$ is the multiplicity for a given sequence
of $(s_1,s_2,\dots )$.  Though it is not explicitly denoted, 
there is a constraint on the values of $s_a$ such that the product of all 
$s_a$'s is consistent with the chiralities of $\psi_n$ and 
$\tilde{\psi}_n$ (which are listed in Table \ref{table2}). 

\medskip 

Similarly, the contribution from the modes $\psi'_n$ and $\tilde{\psi}'_n$ is given by 
\begin{align}
\prod_{n>1}
\prod_{s'_1,s'_2,\dots = \pm 1}
\left(
    1-q^{\omega'_n} {\rm e}^{\frac{i}{2}(s'_1 \phi_1+s'_2 \phi_2 + \dots)}
\right)^{d(s'_1,s'_2,\dots)} \,,
\label{foscp}
\end{align}
where the product of all $s'_a$'s should be consistent with the chiralities of $\psi'_n$ and 
$\tilde{\psi}'_n$ listed in Table \ref{table2}.
Although the expressions given in (\ref{fosc}) and (\ref{foscp}) are rather symbolic, the meaning
of them will be clarified in later explicit evaluations of various amplitudes.

\medskip 

\begin{table}
\begin{center}
\begin{tabular}{c|ccc}
\hline
   & $\gamma^9$ & $\gamma^{1234}$ & $\gamma^{5678}$ \\
\hline
 $\psi_n$, $\chi$ & $-$ & $+$ & $-$ \\
 $\tilde{\psi}_n$ & $+$ & $-$ & $-$ \\
\hline
 $\psi'_n$, $\chi'$ & $-$ & $-$ & $+$ \\
 $\tilde{\psi}'_n$ & $+$ & $+$ & $+$ \\
\hline
\end{tabular}
\end{center}
\caption{\footnotesize The chiralities of the 
fermionic modes for $\gamma^9$, $\gamma^{1234}$, and 
$\gamma^{5678}$.}
\label{table2}
\end{table}

\subsubsection*{The zero modes}

The next is to consider the contributions from the bosonic zero modes. 
For each of the directions without boundary condensates, if it is in the 1234 space (or 
in the 5678 space), the contribution is given by  
\begin{align}
(1-q^{m/3})^{-1/2} \qquad \left(\mbox{or}~~(1-q^{m/6})^{-1/2}\right)\,.
\end{align}

\medskip 

For a two-dimensional plane labeled by $a$ with a boundary condensate, 
if it is in the 1234 space, the contribution is 
\begin{align}
(1-\cos \varphi_a^{(1)} \cos \varphi_a^{(2)} q^{m/3})^{-1}\,,
\end{align}
while, if it is in the 5678 space, the contribution is 
\begin{align}
(1-\cos \varphi_a^{(1)} \cos \varphi_a^{(2)} q^{m/6})^{-1}\,. 
\end{align}

\medskip 

Then let us see the fermionic zero modes. The contribution from the mode $\chi$ is given by 
\begin{align}
\prod_{s_1,s_2,\dots = \pm 1}
\left( 
    1- q^{m/3} {\rm e}^{\frac{i}{2}(s_1 \phi_1+s_2 \phi_2 + \dots)}
\right)^{d(s_1,s_2,\dots)/2}\,, 
\label{fzosc}
\end{align}
where the product of $s_a$'s is under the same constraint given to (\ref{fosc}).  
Similarly, the contribution from the mode $\chi'$ is  
\begin{align}
\prod_{s'_1,s'_2,\dots = \pm 1}
\left( 
    1- q^{m/6} {\rm e}^{\frac{i}{2}(s'_1 \phi_1+s'_2 \phi_2 + \dots)}
\right)^{d(s'_1,s'_2,\dots)/2}\,,  
\label{fzoscp}
\end{align}
where the product of $s'_a$'s is under the same constraint given to
(\ref{foscp}).

\medskip 

By following the general prescription, we will consider concrete examples 
in the next subsection.

\subsection{Examples}

In the following, let us consider three types of D-branes concretely and 
compute the corresponding amplitudes.

\subsubsection*{Parallel D6-branes}

First of all, let us consider parallel D6-branes which extend along the (1,2,4,5,6,7,8) directions 
(see Table \ref{list:tab}). We turn on boundary condensates in the 12, 56, and 78 planes. 
Then the matrix $M$ is expressed as 
\begin{align}
M = \left(
	  \begin{array}{ccccc}
      M_{(1)} &   &    &         &  \\
              &+1 &    &         &   \\
              &   & -1 &         &   \\
              &   &    & M_{(2)} &    \\
              &   &    &         & M_{(3)}
      \end{array}
    \right)\,.
\end{align}
Here $M_{(a)}={\rm e}^{\varphi_a T_{(a)}}$ $(a=1,2,3)$ is the $2\times 2$
block given in (\ref{bmblock})\,, where $T_{(1)}$, $T_{(2)}$, and $T_{(3)}$ 
are the rotation generators in the 12, 56, and 78 planes, respectively.  

\medskip 

On the other hand, $\hat{M}$ is given by 
\begin{align}
\hat{M} = {\rm e}^{\frac{\varphi_1}{2} \gamma^{12}} \gamma^4
          {\rm e}^{\frac{\varphi_2}{2} \gamma^{56}} 
          {\rm e}^{\frac{\varphi_3}{2} \gamma^{78}} 
        = {\rm e}^{\frac{\varphi_1}{2} \gamma^{12}}
          {\rm e}^{\frac{\varphi_2}{2} \gamma^{56}} 
          {\rm e}^{\frac{\varphi_3}{2} \gamma^{78}} \gamma^4 \,.
\end{align}
For the contributions of $\psi_n$, $\tilde{\psi}_n$, and $\chi$ to
the amplitude, the
sequences of eigenvalues $(s_1,s_2,s_3)$ in (\ref{fosc}) and 
(\ref{fzosc}) are obtained from Table \ref{table2} as
\begin{align}
(s_1,s_2,s_3) = (\pm, +, +) \,, \, (\pm,-,-)\,,
\end{align} 
for each of which the multiplicity is 1, $d(s_1,s_2,s_3)=1$\,.
For $\psi'_n$, $\tilde{\psi}'_n$, and $\chi'$, the
sequences of eigenvalues $(s'_1,s'_2,s'_3)$ are 
\begin{align}
(s'_1,s'_2,s'_3) = (\pm, +, -) \,, \, (\pm, -, +)\,,  
\end{align}
for each of which the multiplicity is 1.

\medskip 

The basic building blocks from (\ref{builds}) to 
(\ref{fzoscp}) lead to the following factorized form, 
\begin{align}
\tilde{\mathcal{A}}_{{\rm D}6;{\rm D}6} (t) =
\tilde{\mathcal{A}}^{(0)}_{{\rm D}6;{\rm D}6} (t)
\tilde{\mathcal{A}}^{(\text{osc})}_{{\rm D}6;{\rm D}6} (t)\,.
\end{align}
Here $\tilde{\mathcal{A}}^{(0)}_{D6;D6} (t)$ comes from the zero mode, 
\begin{align}
\tilde{\mathcal{A}}^{(0)}_{{\rm D}6;{\rm D}6} (t) 
=& (1-q^{m/3})^{-1}
   (1-\cos \varphi_1^{(1)} \cos \varphi_1^{(2)} q^{m/3})^{-1}
\notag \\
& \times 
   (1-\cos \varphi_2^{(1)} \cos \varphi_2^{(2)} q^{m/6})^{-1}
   (1-\cos \varphi_3^{(1)} \cos \varphi_3^{(2)} q^{m/6})^{-1}
\notag \\
& \times
  \left( 
    1- q^{m/3} {\rm e}^{\frac{i}{2}(\phi_1+ \phi_2 + \phi_3)}
  \right)^{1/2}
  \left( 
    1- q^{m/3} {\rm e}^{\frac{i}{2}(-\phi_1+ \phi_2 + \phi_3)}
  \right)^{1/2}
\notag \\
& \times
  \left( 
    1- q^{m/3} {\rm e}^{\frac{i}{2}(\phi_1 - \phi_2 - \phi_3)}
  \right)^{1/2}
  \left( 
    1- q^{m/3} {\rm e}^{\frac{i}{2}(-\phi_1 - \phi_2 - \phi_3)}
  \right)^{1/2}
\notag \\
& \times
  \left( 
    1- q^{m/6} {\rm e}^{\frac{i}{2}(\phi_1+ \phi_2 - \phi_3)}
  \right)^{1/2}
  \left( 
    1- q^{m/6} {\rm e}^{\frac{i}{2}(-\phi_1+ \phi_2 - \phi_3)}
  \right)^{1/2}
\notag \\
& \times
  \left( 
    1- q^{m/6} {\rm e}^{\frac{i}{2}(\phi_1 - \phi_2 + \phi_3)}
  \right)^{1/2}
  \left( 
    1- q^{m/6} {\rm e}^{\frac{i}{2}(-\phi_1 - \phi_2 + \phi_3)}
  \right)^{1/2}\,,
\end{align}
and $\tilde{\mathcal{A}}^{(\text{osc})}_{D6;D6} (t)$ is from the nonzero mode, 
\begin{align}
\tilde{\mathcal{A}}^{(\text{osc})}_{{\rm D}6;{\rm D}6} (t)
=& \prod_{n>0}(1-q^{\omega_n})^{-2}
     (1-q^{\omega_n} {\rm e}^{i \phi_1})^{-1}
     (1-q^{\omega_n} {\rm e}^{-i \phi_1})^{-1}
\notag \\
& \times
     (1-q^{\omega'_n} {\rm e}^{i \phi_2})^{-1}
     (1-q^{\omega'_n} {\rm e}^{-i \phi_2})^{-1}
     (1-q^{\omega'_n} {\rm e}^{i \phi_3})^{-1}
     (1-q^{\omega'_n} {\rm e}^{-i \phi_3})^{-1}
\notag \\
& \times
  \left( 
    1- q^{\omega_n} {\rm e}^{\frac{i}{2}(\phi_1+ \phi_2 + \phi_3)}
  \right)
  \left( 
    1- q^{\omega_n} {\rm e}^{\frac{i}{2}(-\phi_1+ \phi_2 + \phi_3)}
  \right)
\notag \\
& \times
  \left( 
    1- q^{\omega_n} {\rm e}^{\frac{i}{2}(\phi_1 - \phi_2 - \phi_3)}
  \right)
  \left( 
    1- q^{\omega_n} {\rm e}^{\frac{i}{2}(-\phi_1 - \phi_2 - \phi_3)}
  \right)
\notag \\
& \times
  \left( 
    1- q^{\omega'_n} {\rm e}^{\frac{i}{2}(\phi_1+ \phi_2 - \phi_3)}
  \right)
  \left( 
    1- q^{\omega'_n} {\rm e}^{\frac{i}{2}(-\phi_1+ \phi_2 - \phi_3)}
  \right)
\notag \\
& \times
  \left( 
    1- q^{\omega'_n} {\rm e}^{\frac{i}{2}(\phi_1 - \phi_2 + \phi_3)}
  \right)
  \left( 
    1- q^{\omega'_n} {\rm e}^{\frac{i}{2}(-\phi_1 - \phi_2 + \phi_3)}
  \right)\,. 
\end{align} 

It is now convenient to introduce the ``massive'' thetalike function defined in
\cite{Takayanagi:2002pi,Sugawara:2002rs}
\begin{align}
\Theta_{(a,b)} (\tau; \nu) \equiv
e^{4\pi \tau_2 \Delta(\nu; a) }
\prod_{n \in \mathbf{Z}} 
\left|
    1 - {\rm e}^{-2\pi \tau_2 \sqrt{\nu^2 + (n+a)^2} + 2\pi \tau_1 (n+a)
            +2\pi i b }
\right|^2\,, 
\label{thetafn}
\end{align}
where $\tau = \tau_1 + i \tau_2$ and $\Delta(\nu;a)$ is the
zero-point energy. Actually, we are not concerned with $\Delta(\nu;a)$ 
because it disappears in the final expression. 

\medskip 

By using the massive theta functions, the amplitude can be rewritten 
into a simpler form,  
\begin{align}
\tilde{\mathcal{A}}_{{\rm D}6;{\rm D}6} (t) 
= &  
 \frac{(1-q^{m/3}{\rm e}^{i\phi_1})^{1/2} (1-q^{m/3}{\rm e}^{-i\phi_1})^{1/2}}
        {(1-\cos \varphi_1^{(1)} \cos \varphi_1^{(2)} q^{m/3} )}
   \times
 \frac{(1-q^{m/6}{\rm e}^{i\phi_2})^{1/2} (1-q^{m/6}{\rm e}^{-i\phi_2})^{1/2}}
        {(1-\cos \varphi_2^{(1)} \cos \varphi_2^{(2)} q^{m/6} )}
\notag \\
& \times
  \frac{(1-q^{m/6}{\rm e}^{i\phi_3})^{1/2} (1-q^{m/6}{\rm e}^{-i\phi_3})^{1/2}}
        {(1-\cos \varphi_3^{(1)} \cos \varphi_3^{(2)} q^{m/6} )} 
\notag \\
& \times
\frac{
      \Theta^{1/2}_{(0,(\phi_1+\phi_2+\phi_3)/4\pi)}(it;m/3)
      \Theta^{1/2}_{(0,(\phi_1-\phi_2-\phi_3)/4\pi)}(it;m/3)
      }
      {
      \Theta^{1/2}_{(0,0)}(it;m/3)
	  \Theta^{1/2}_{(0,\phi_1 /2\pi)}(it;m/3)
	  }
\notag \\
& \times
\frac{
      \Theta^{1/2}_{(0,(\phi_1+\phi_2-\phi_3)/4\pi)}(it;m/6)
      \Theta^{1/2}_{(0,(\phi_1-\phi_2+\phi_3)/4\pi)}(it;m/6)
      }
      {
      \Theta^{1/2}_{(0,\phi_2 /2\pi)}(it;m/6)
      \Theta^{1/2}_{(0,\phi_3 /2\pi)}(it;m/6)
	  }\,.
\label{d6amp}
\end{align}
We would like to note that, if the boundary condensates are absent,
the above amplitude simply becomes one.  This is consistent with
the previous result \cite{Shin:2003ae} of the amplitude calculation
without boundary condensates.

\subsubsection*{Parallel D4-branes}

Let us consider parallel D4-branes.  
The world volume of the D4-branes extends along the (1,2,3,5,6) directions. 
Then we turn on boundary condensates in the 1-2 and 5-6 planes.  

\medskip 

It is possible to compute the amplitude in the same way as in the case of D6-branes.  
But it can also be obtained simply by taking the magnetic
background in the 78 plane to be infinite from the result on D6-branes. 
Note that $x^3$ ($x^4$) should be understood as the Neumann (Dirichlet) direction at that time.

\medskip 

Let us consider the limit $f^{(1)}_3, f^{(2)}_3 \rightarrow \infty$ (equivalently 
$\varphi^{(1)}_3, \varphi^{(2)}_3 \rightarrow 0$, and thus
$\phi_3 \rightarrow 0$) in (\ref{d6amp})\,. Then the $x^7$ and $x^8$ directions are turned  
into the Dirichlet ones.  The resulting amplitude is given by 
\begin{align}
\tilde{\mathcal{A}}_{{\rm D}4;{\rm D}4} (t) 
=&   
 \frac{(1-q^{m/3}{\rm e}^{i\phi_1})^{1/2} (1-q^{m/3}{\rm e}^{-i\phi_1})^{1/2}}
        {(1-\cos \varphi_1^{(1)} \cos \varphi_1^{(2)} q^{m/3} )}
   \times
 \frac{(1-q^{m/6}{\rm e}^{i\phi_2})^{1/2} (1-q^{m/6}{\rm e}^{-i\phi_2})^{1/2}}
        {(1-\cos \varphi_2^{(1)} \cos \varphi_2^{(2)} q^{m/6} )}
\notag \\
& \times
\frac{
      \Theta^{1/2}_{(0,(\phi_1+\phi_2)/4\pi)}(it;m/3)
      \Theta^{1/2}_{(0,(\phi_1-\phi_2)/4\pi)}(it;m/3)
      }
      {
      \Theta^{1/2}_{(0,0)}(it;m/3)
	  \Theta^{1/2}_{(0,\phi_1 /2\pi)}(it;m/3)
	  }
\notag \\
& \times
\frac{
       \Theta^{1/2}_{(0,(\phi_1+\phi_2)/4\pi)}(it;m/6)
      \Theta^{1/2}_{(0,(\phi_1-\phi_2)/4\pi)}(it;m/6)
      }
      {
	  \Theta^{1/2}_{(0,0)}(it;m/6)
	  \Theta^{1/2}_{(0,\phi_2 /2\pi)}(it;m/6)
	  }\,.
\label{d4amp}
\end{align}

\subsubsection*{Parallel D2-branes}

Finally, we shall consider parallel D2-branes. 
The world-volume of the D2 branes expands along the $(1,2,4)$ directions. 
We turn on the boundary condensate in the 1-2 plane.  
Then it is straightforward to obtain the following amplitude, 
\begin{align}
\tilde{\mathcal{A}}_{{\rm D}2;{\rm D}2} (t)
=& \frac{ (1-2 \cos (\phi/2) q^{m/3} + q^{2m/3})
          (1-2 \cos (\phi/2) q^{m/6} + q^{2m/6}) 
        }{ (1-q^{m/3}) (1-q^{m/6})^2 
           (1-\cos \varphi^{(1)} \cos \varphi^{(2)} q^{m/3} )
        } 
\notag \\
& \times \prod_{n>0}
  \frac{ (1-2 \cos (\phi/2) q^{\omega_n} + q^{2\omega_n})^2
          (1-2 \cos (\phi/2) q^{\omega'_n} + q^{2\omega'_n})^2 
        }{ (1-q^{\omega_n})^2 (1-q^{\omega'_n})^4 
           (1-2 \cos (\phi) q^{\omega_n} + q^{2\omega_n})
        }\,. 
\end{align}
This is also obtained from the amplitude for the D4-branes (\ref{d4amp})
by taking $f^{(1)}_2, f^{(2)}_2 \rightarrow \infty$ (equivalently 
$\varphi^{(1)}_2, \varphi^{(2)}_2 \rightarrow 0$, and thus
$\phi_2 \rightarrow 0$)\,. 

\medskip 

In terms of the massive thetalike function, the amplitude can be rewritten as 
\begin{align}
\tilde{\mathcal{A}}_{{\rm D}2;{\rm D}2} (t)
=& \frac{(1-q^{m/3}{\rm e}^{i\phi})^{1/2} (1-q^{m/3}{\rm e}^{-i\phi})^{1/2}}
        {(1-\cos \varphi^{(1)} \cos \varphi^{(2)} q^{m/3} )}
\notag \\
&\times \frac{ \Theta_{(0,\phi/4\pi)}(it; m/3) 
          \Theta_{(0,\phi/4\pi)}(it; m/6)}
        { \Theta_{(0,0)}^{1/2}(it; m/3)
          \Theta_{(0,\phi/2\pi)}^{1/2} (it; m/3)
          \Theta_{(0,0)}(it; m/6)
        }\,, 
\label{d2amp}
\end{align}
where the subscript 1 has been omitted from all of the angles because
the boundary condensate is turned on only in the 1-2 plane.

\section{No pair production of open strings}

This section considers the possibility of the pair production of open strings 
from the viewpoint of the pole structure of the amplitudes. 
As an example, we concentrate on the case of two parallel D2-branes.   

\medskip 

First of all, it is necessary to move from the cylinder diagram 
(closed string channel) to an annulus one (open string channel). 
It is carried out by performing the transformation 
$t \rightarrow t'=1/t$\,. 
By the way, the resulting amplitude is the one containing the magnetic
condensate.  In order to investigate the issue of open string pair
production, it is necessary to have an electric condensate. 
It is difficult to accomplish it directly because we are now working 
with the light-cone gauge.  Still, the amplitude with an electric
condensate can be anticipated from the magnetic one through the 
replacement $f\rightarrow if$ 
(equivalently $\phi \rightarrow i \phi$)\,\footnote{The replacement is 
not to be regarded as the result of signature change from the Euclidean
space to the Minkowskian one, because the time direction is definitely 
set by the light-cone time in our computation.  Rather, it should be
anticipated as a natural generalization of the resulting amplitude
following the reasoning of \cite{Tseytlin}. For the issue of signature
change from a rigorous viewpoint, see for example 
Ref.~\cite{Bachas:2003sj}.}. 
This anticipation will be supported later from agreement with the result 
in the flat limit.

\medskip 

Then the pole structure on the real $t'$ axis leads to an imaginary
part in the expression of the energy, which is given by the sum over 
the residues of the poles. This is interpreted as the sign of the pair creation of open strings. 

\medskip 

To see the pole structure, recall the transformation law of massive 
thetalike functions
under the $S$ transformation $\tau \rightarrow -1/\tau$
\cite{Takayanagi:2002pi,Sugawara:2002rs}\,,
\begin{align}
\Theta_{(a,b)} \left(-\frac{1}{\tau}; |\tau|\nu\right)
= \Theta_{(b,-a)} (\tau; \nu)\,.
\end{align}
Then the D2 amplitude in (\ref{d2amp}) is rewritten as 
\begin{eqnarray}
\tilde{\mathcal{A}}_{{\rm D}2;{\rm D}2} (t)
&=& \frac{\Theta_{(0,\phi/4\pi)}(it; m/3) 
          \Theta_{(0,\phi/4\pi)}(it; m/6)
          }
          {\Theta^{1/2}_{(0,\phi/2 \pi)} (it; m/3)
          } \times \dots
\notag \\
&\stackrel{t\rightarrow t'=1/t}{\longrightarrow} &
\frac{\Theta_{(-\phi/4\pi,0)}(it'; m/3t') 
      \Theta_{(-\phi/4\pi,0)}(it'; m/6t')
      }
      {\Theta^{1/2}_{(-\phi/2 \pi,0)} (it'; m/3t')
      } \times \dots
\notag \\
&\stackrel{\phi \rightarrow i \phi}{\longrightarrow} &
\frac{\Theta_{(-i\phi/4\pi,0)}(it'; m/3t') 
      \Theta_{(-i\phi/4\pi,0)}(it'; m/6t')
      }
      {\Theta^{1/2}_{(-i\phi/2 \pi,0)} (it'; m/3t')
      } \times \dots
\label{pair-amp}
\end{eqnarray}
where ``...'' denote the factors irrelevant to the pole structure. 
By the definition of the massive thetalike function
(\ref{thetafn})\,, the last line does not lead to any pole in the real $t'$ axis 
(For details of the proof, see Appendix A). 
Thus it has been shown that there is no pair creation of open strings.

\medskip 

One interpretation of this result is that strings are trapped in a harmonic potential
due to the IIA plane wave background.  In other words, the string
coordinates describe the set of harmonic oscillators. Hence 
it is impossible to separate the constituents of the produced pairs in an infinite
distance even if a pair is produced at a certain instance.

\medskip

At this point, one may ask if the production rate becomes finite 
after removing the harmonic potential, that is, after taking the
flat space-time limit ($m \rightarrow 0$)\,.  If it tends to be finite, 
then the well-known result in flat space-time is reproduced
and our interpretation passes an important check.

\medskip 

Let us consider the flat space-time limit. 
With the help of the expression of the massive
thetalike function in $m \rightarrow 0$ limit \cite{Sugawara:2002rs},
\begin{align}
\lim_{m \rightarrow 0} \Theta_{(a,b)} (\tau ; m) = {\rm e}^{-2\pi \tau_2 a^2} 
\left| \frac{\theta_1 (a\tau + b |\tau )}{\eta(\tau)} \right|^2
\end{align}
and the usual product form of Jacobi theta function,
\begin{align}
\theta_1(z|\tau) 
 = 2 q^{\frac{1}{8}} \sin ( \pi z )
   \prod^\infty_{n=1} (1-q^n) (1-q^n {\rm e}^{2 \pi i z} )
                        (1-q^n {\rm e}^{-2 \pi i z} ) 
\qquad (q \equiv {\rm e}^{2\pi i \tau}) \,,
\end{align}
the last line of (\ref{pair-amp}) can be rewritten as 
\begin{eqnarray}
&& \frac{\Theta_{(-i\phi/4\pi,0)}(it'; m/3t') 
      \Theta_{(-i\phi/4\pi,0)}(it'; m/6t')
      }
      {\Theta^{1/2}_{(-i\phi/2 \pi,0)} (it'; m/3t')
      } \times \dots  \notag \\ 
& \stackrel{ m \rightarrow 0 }{\longrightarrow} &  \qquad
  \frac{\theta^4_1 \left( \frac{\phi t'}{4\pi} \big| it' \right) }
       { \theta_1 \left( \frac{\phi t'}{2\pi} \big| it' \right)} 
  \times \dots
\qquad \longrightarrow \qquad
  \frac{ \sin^4 \left( \frac{\phi t'}{4} \right) }
       { \sin \left( \frac{\phi t'}{2} \right) }
  \times \dots
\end{eqnarray}
The last line explicitly shows that there are an infinite number
of poles on the real $t'$ axis. The locations of the poles are specified by  
\[
t' = 2\pi (2k +1)/\phi \qquad (k \in \mathbf{Z})\,.
\]  
This is nothing but the result in flat spacetime \cite{Green:1996um}\footnote{  
Although D-strings in type IIB string theory are considered there, the essential point is the same.}.

\section{Conclusion and Discussion}
\label{concl}

We have considered whether an external electric field may cause the pair production of open strings 
in a type IIA plane-wave background. 
The boundary states of D-branes with condensates have been constructed 
in the Green-Schwarz formulation with the light-cone gauge. 
The cylinder diagrams have been computed with the boundary states and 
the resulting amplitudes are shown to be expressed in terms of massive theta functions. 
This is a characteristic property intrinsic to plane-wave backgrounds. 
As a consequence,  although the value of the electric field is bounded by the upper value,\footnote{We note that the upper bound is not the usual
constant critical electric field and comes from the
consideration of Eq. (\ref{1}), the zero-mode part that is essential
in investigating the pair production.  From the viewpoint of $\phi$, it 
is easy to see that $\phi$ has the upper bound from the non-negativity 
of the inside of the square root, which is given by (\ref{2}).} 
there is no pole in the amplitudes and it indicates that no pair 
production occurs in the plane-wave background. 
Our result is based on an analysis in a IIA pp-wave background, but 
the result would be universal for a class of plane-wave backgrounds.  

\medskip 

In order to confirm our conjecture that no pair production occurs, 
it is indispensable to compute the amplitudes in other plane-wave backgrounds. 
It would be interesting to classify the 
gravitational backgrounds which allows the pair production. 
For example, plane-wave backgrounds with flat directions are good 
candidates. In this sense, adding angular momenta would be
able to support the pair production. 

\medskip 

The next important question is whether or not the result of no pair
production is intrinsic to plane-wave backgrounds.
As was stated in the Introduction,
it is interesting to study the possibility of the pair production in AdS backgrounds.
A plane-wave background appears as an approximation of the AdS geometry times an internal space, 
while the geometry of flat space always appears by considering a small and local region 
and the pair creation seems possible on it.  
Actually, a Penrose limit \cite{Penrose} of the AdS geometry 
may lead to flat space, depending on the choice of the null 
geodesic. Thus, our argument would not be able to exclude 
the possibility that the pair production of strings occurs in the AdS backgrounds. 
The phenomenon that no pair production occurs may be an artifact in the plane-wave approximation.

\medskip 

The study of the pair production of open strings on curved backgrounds 
would reveal a new aspect of the string dynamics.

\section*{Acknowledgments}

K.Y. would like to thank T.~Enari and A.~Miwa for discussions.  
H.S. would like to thank F.~Sugino for discussions and acknowledges 
the hospitality of the Okayama Institute for Quantum Physics (OIQP), where 
this work was completed during his visit.
The work of H.S. was supported by the National Research Foundation of 
Korea (NRF) grant funded by the Korea government (MEST)  with Grants
No.~NRF-2012R1A1A2004203, No.~2012-009117, and No.~2012-046278.

\appendix

\section*{Appendix}

\section{The pole structure of the D2-brane amplitude}
\label{app1}

We will show that there is no pole in the D2-brane amplitude given in (\ref{pair-amp})\,. 

\medskip 

Let us begin with the last line of (\ref{pair-amp})\,. The definition of the massive theta function is given in (\ref{thetafn}).
The goal is to show that there is no pole on the real $t'$ axis. The amplitude is divided into 
(1) the $n \neq 0$ contributions and (2) the $n=0$ contribution. We will consider each of them below.

\subsection{The $n \neq 0$ contributions} 

In order to study the pole structure, let us consider zero points of the following part, 
\begin{eqnarray}
\Theta_{(-i\phi/2\pi,0)} \left(it';\frac{m}{3t'}\right) 
= {\rm e}^{4\pi t'\Delta(m/3t'; -i\phi/2\pi)} \prod_{n\in \mathbb{Z}} 
\left| 1- \exp\left[
-\sqrt{ M^2 + t'{}^2(2\pi n - i\phi)^2}\,\right] \right|^2\,, 
\end{eqnarray}
where we have defined as 
\[
M \equiv \frac{2\pi m}{3}\,.
\]
It is convenient to rewrite the argument of the exponential part as follows: 
\begin{eqnarray} 
M^2 +t'{}^2(2\pi n -i\phi)^2 
&=& M^2 + t'{}^2\left[(2\pi n)^2 - \phi^2\right] - i \cdot 4\pi n t'{}^2\phi \nonumber \\ 
&\equiv & r {\rm e}^{i\varphi} = r \cos\varphi + i r\sin \varphi\,. 
\end{eqnarray}
Here the parameters are identified as 
\begin{eqnarray}
r \sin\varphi = -4\pi n t'{}^2 \phi\,, \qquad r \cos\varphi = M^2 +t'{}^2\left[(2\pi n)^2 - \phi^2\right]\,,
\end{eqnarray}
where $r$ and $\varphi$ are represented by 
\begin{eqnarray}
r^2 &=& (4\pi n t'{}^2\phi)^2 + \left(M^2 + t'{}^2[(2\pi n)^2 - \phi^2] \right)^2\,,  \nonumber \\ 
\tan\varphi &=& \frac{-4\pi n t'{}^2\phi}{M^2 + t'{}^2[(2\pi n)^2 - \phi^2]}\,.  \label{const}
\end{eqnarray}
Then it is easy to derive the following expression, 
\begin{eqnarray}
\left| 1-\exp\left[
-\sqrt{ M^2 + t'{}^2(2\pi n - i\phi)^2}\,
\right] \right|^2 
= 2- 2{\rm e}^{-r^{1/2}\cos(\varphi/2)}\cos\left(r^{1/2}\sin(\varphi/2)\right)\,. 
\end{eqnarray}
Because $r \neq 0$ on the real $t'$ axis, 
the only condition that poles exist is the following,  
\[
\varphi=\pi\,, \qquad r^{1/2} = 2\pi N \quad (N\in \mathbb{N})\,.
\] 
However, this condition is not satisfied due to the condition (\ref{const})\,.

\subsection{The $n=0$ contribution}

The next is to see the contribution from the $n=0$ mode. 

\medskip 

From the denominator of (\ref{pair-amp})\,, one can read off the $n=0$ contribution as follows: 
\begin{align}
\Theta^{1/2}_{(-i\phi/2 \pi,0)} (it'; m/3t') \quad \longrightarrow \quad
1- \exp\left[-2\pi t' 
    \sqrt{\left( \frac{m}{3t'} \right)^2 
           - \left( \frac{\phi}{2\pi} \right)^2 }\,\right]\,. 
\label{1}
\end{align}
It is easy to see that this factor becomes 0 at 
\begin{align}
t' = \frac{m}{3} \cdot \frac{2\pi}{\phi} \,. 
\label{2}
\end{align}
At this point, it seems that there should be a pole at this value. 

\medskip

On the other hand, a massive thetalike function on the numerator of (\ref{pair-amp}) 
contains the $n=0$ contribution given by 
\begin{align}
\Theta_{(-i\phi/4 \pi,0)} (it'; m/6t') \quad \longrightarrow \quad
\left( 1- \exp\left[-2\pi t' 
    \sqrt{\left( \frac{m}{6t'} \right)^2 
           - \left( \frac{\phi}{4\pi} \right)^2 }\,\right] \right)^2 \,. \label{3}
\end{align}
Interestingly, this factor also becomes 0 at the value of $t'$ in (\ref{2}).
Noting that the power of (\ref{3}) is higher than that of (\ref{1})\,, 
the value of $t'$ in (\ref{2}) does not indicate the existence of a pole 
but a vanishing point of the amplitude.

\medskip 

In total, we conclude that the D2-brane amplitude given in (5.2) does not have any pole
on the real $t'$ axis.

\end{document}